\documentclass[aps,prb,twocolumn,superscriptaddress,print]{revtex4-2}
\usepackage{amssymb,amsmath}
\usepackage{graphicx}
\usepackage{multirow}
\usepackage{hyperref}
\usepackage{appendix}
\usepackage{mathtools}
\usepackage{amsmath}
\usepackage{verbatim}
\usepackage{gensymb}
\usepackage{subfigure}
\usepackage{bm}
\usepackage{threeparttable}
\usepackage{booktabs}
\usepackage{braket}
\usepackage{xcolor}
\usepackage[normalem]{ulem}
\usepackage[mathscr]{euscript}
\usepackage{tabularx}

\begin{document}

\title{Tunable periodic cascade of delocalization transitions in non-Hermitian IAAF model}%
\author{Yi Xue}
\affiliation{Lanzhou Center for Theoretical Physics, Key Laboratory of Quantum Theory and Applications of the Ministry of Education, Key Laboratory of Theoretical Physics of Gansu Province, Gansu Provincial Research Center for Basic Disciplines of Quantum Physics, School of Physics and Technology, Lanzhou University, Lanzhou 730000, China}

\author{Yunhua Wang}
\email{wangyunhua@lzu.edu.cn}
\affiliation{Lanzhou Center for Theoretical Physics, Key Laboratory of Quantum Theory and Applications of the Ministry of Education, Key Laboratory of Theoretical Physics of Gansu Province, Gansu Provincial Research Center for Basic Disciplines of Quantum Physics, School of Physics and Technology, Lanzhou University, Lanzhou 730000, China}

\begin{abstract}
Cascade of delocalization transitions provides a route to multi-step control of delocalization, yet achieving a programmable periodic cascade remains a challenge. Here, we demonstrate a tunable periodic cascade of delocalization transitions in a non-Hermitian interpolating Aubry–André–Fibonacci (IAAF) model by introducing an imaginary phase shift into the quasiperiodic potential. Such a periodic cascade of delocalization transitions arises from the competition between kinetic-energy-facilitated delocalization and gain-loss-regulated localization. In numerical static localization characterizations, successive regular stripe patterns for localized and extended states emerge in phase diagrams. We also derive an analytical expression for the cascade period and find that it relies solely on the imaginary phase shift. This provides a direct way of continuously tuning the cascade period via a single control parameter. Complementary static and dynamical diagnostics consistently confirm this periodic cascade of delocalization transitions in the non-Hermitian IAAF system. Our findings identify a novel class of delocalization transitions with tunable cascading periods, opening a route toward experimentally programmable control of delocalization. 
\end{abstract}

\maketitle

\section{Introduction}
Cascades of delocalization transitions refer to a new delocalization transition phenomenon, in which, as a control parameter is varied continuously, eigenstates repeatedly switch between localized and extended phases, giving rise to a succession of delocalization and relocalization transitions. In 2020, Goblot et al. predicted such a cascade theoretically in the interpolating Aubry–André–Fibonacci (IAAF) model and observed its transition in a cavity-polariton experiment \cite{Goblot2020}. This established that localizations can be controlled not only through a single critical point, but through a sequence of transitions. Since then, cascaded and reentrant localization transitions have been predicted or realized in Aubry–André–Harper (AAH) and IAAF-type quasiperiodic systems by modifying the potential, introducing staggered structures or additional hopping, applying time-periodic driving, or enhancing off-diagonal modulation \cite{Roy2021,Padhan2022,Li2023,Wei2025,Dai2023,Wang2023_PRB107,Zhou2022_PRB106,Aditya2023_PRB107,Zhou2021,Tabanelli2024,Chang2026}. These developments have established the cascade of delocalization transitions as a route toward multi-step control of localizations.

Non-Hermiticity provides another powerful mechanism for controlling localizations through non-reciprocal hopping, gain and loss, and complex potentials \cite{Hatano1996,Ashida2020,Bergholtz2021,ElGanainy2018}. When combined with quasiperiodicity, these ingredients can generate additional localization and delocalization transitions. Examples include the IAAF model with non-reciprocal hopping \cite{Zhai2021} and AAH-type chains with an imaginary phase shift in the quasiperiodic potential \cite{Longhi2019,Liu2020,Liu2021}. Such localization phenomena have been explored in cold atoms and optical lattices, photonic lattices and waveguides, cavity polaritons, quantum walks, and electrical circuits \cite{Roati2008,Lahini2009,Goblot2020,Lin2022,Su2023_exp,Vaidya2022,Weidemann2022}, demonstrating the broad experimental accessibility of quasiperiodic localization transitions.

Despite these advances, existing cascaded or reentrant transitions generally arise from the interplay of several different mechanisms, and their transition points are obtained through complex numerical computations rather than analytical evaluations. Moreover, the cascaded or reentrant transitions in parameter space are irregular \cite{Roy2021,Zhai2021,Padhan2022,Li2023,Wei2025,Guo2024,Han2022_PRB105,Jiang2021_CPB,Wu2021_NJP,Wang2023_PRB107}. It is hoped that one tunable periodic cascade of delocalization transitions can be realized in an experimentally accessible way. This naturally motivates a question: how can one realize a tunable periodic cascade of delocalization transitions in which the sequence of localization changes is periodic, its spacing is analytically predictable, and its period can be continuously tuned by a single control parameter.

Herein, we show that such a tunable periodic cascade of delocalization transitions can be realized in a non-Hermitian IAAF model with an imaginary phase shift characterizing the non-Hermiticity \cite{Longhi2019,Liu2020,Liu2021}. Such a periodic cascade of delocalization transitions under a moderate potential magnitude in this system is attributed to the competition between kinetic-energy-facilitated delocalization and gain-loss-regulated localization. We evaluate an analytical formula for the cascade period of these transitions and find the cascade period depends solely on the imaginary phase shift with a tunable feature, i.e., increasing the imaginary phase shift leading to a reduction of the cascade period. Numerical computation results of the static localization corroborate both the periodic cascading delocalization-transition phenomenon and analytical predictions.

We further investigate short-time and long-time dynamical characteristics of wave-packet expansion across delocalization transitions using time-dependent root-mean-square displacements and wave-spreading profiles \cite{Ketzmerick1997,Thiem2009,Orito2022}. The short-time dynamical calculations validate the periodic cascade of delocalization transitions in this non-Hermitian IAAF system. Moreover, wave-packet expansion exhibits pseudo-localization under the pole condition \cite{Longhi2022}. Over sufficiently long time scales, all complex-energy states asymptotically settle into pseudo-localized states, originating from the exponential growth of wavefunctions associated with complex energies. When the potential magnitude becomes relatively larger, localization prevails instead since extended states are suppressed greatly. Our work identifies a novel class of delocalization transitions with tunable cascading periods and offers a pathway toward experimentally realizing periodic cascades of delocalization transitions.

\section{Non-Hermitian IAAF model}
We consider a non-Hermitian IAAF model with its Hamiltonian reading
\begin{equation}
H = \sum_{k=1}^{L-1} t(c_k^\dagger c_{k+1} + c^\dagger_{k+1}c_k) + \sum_{k=1}^{L} V_k(\beta) c_k^\dagger c_k,
\label{eq:hamiltonian}
\end{equation}
where $c_k^\dagger$ ($c_k$) are the creation (annihilation) operators at site $k$, $t$ is the hopping amplitude and is set as the energy unit, i.e., $t=1$, and $L$ is the total number of sites. Different from the non-Hermiticity induced by non-reciprocal hopping in Refs. \cite{Hatano1996,Jiang2019}, here in Eq.~\eqref{eq:hamiltonian} the non-Hermiticity arises from the complex on-site potential $V_k(\beta)$ defined as
\begin{equation}
V_k(\beta) = -\lambda \frac{\tanh(\beta \varphi_k)}{\tanh(\beta)},
\label{eq:potential}
\end{equation}
with 
\begin{equation}
\varphi_k = \cos(2\pi\alpha k + i\gamma) - \cos(\pi\alpha),
\label{eq:phi}
\end{equation}
where $\alpha = (\sqrt{5}-1)/2$ is the spatial modulation frequency reflecting the incommensurability of the system, $\lambda$ is the potential strength, $\beta$ is a tunable parameter, and nonzero $\gamma$ characterizes the non-Hermiticity of the system. In Eq.~\eqref{eq:potential}, if $\beta \to 0$, $V_k \to -\lambda\varphi_k$, i.e., the potential reduces to the non-Hermitian AA model. If $\beta \to \infty$, $V_k \to \pm \lambda$, i.e., the potential reduces to the Fibonacci sequence \cite{Jagannathan2021}. If $\gamma = 0$, the potential $V_k(\beta)$ represents the Hermitian IAAF model \cite{Goblot2020,Guo2024}.

The non-Hermiticity induced by complex potential has been extensively studied in theory and experiment in the past few decades \cite{ElGanainy2018,Kawabata2019,Bender1998,Bender2007,Ruter2010,Zhen2015}. The complex potential in Eq.~\eqref{eq:potential} is expressed by the imaginary phase $i\gamma$ in Eq.~\eqref{eq:phi}. Consequently, the tunable parameter $\gamma$ is used to change the real and imaginary parts of the potential. We now discuss the periodic property and singular behavior of the IAAF potential $V_k(\beta)$ in Eq.~\eqref{eq:potential} in the complex plane and associated physics. First, $V_k(\beta)$ has an imaginary period $i\pi$, due to $\tanh(\beta \varphi_k) = \tanh(\beta \varphi_k +i\pi)$. Second, due to $\tanh(\beta \varphi_k) = \sinh(\beta \varphi_k)/\cosh(\beta \varphi_k)$, $V_k(\beta)$ has poles on the imaginary axis, 
\begin{equation}
\begin{split}
\beta\varphi_k =i  \frac{(2n+1)\pi}{2}, \quad n = 0, \pm1, \pm2, \ldots.   
\end{split}
\label{eq:poles}
\end{equation}
Near these poles in Eq.~\eqref{eq:poles}, the IAAF potential $V_k(\beta)$ has a finite real part and a very large imaginary part. The interplay between the periodic property and large imaginary part of $V_k(\beta)$ close to poles may render strong attenuation and localization of wave functions in loss and gain regions, respectively. Consequently, it is natural to expect that the competition between kinetic-energy-driven delocalization and gain-loss-modulated localization enables the tunable periodic cascade of delocalization transitions to be realized in non-Hermitian IAAF model.

\section{Cascade period $\Delta\beta$ of delocalization transitions}
For the IAAF potential $V_k(\beta)$, the localization phase diagram can be mapped in the tunable parameter space ($\beta$, $\lambda$). According to Eq.~\eqref{eq:poles}, the period relies on $\beta$ rather than $\lambda$. Therefore, we characterize the cascade period of delocalization transitions by the period of the tunable parameter $\beta$, namely, $\Delta\beta$. We rewrite Eq.~\eqref{eq:phi} as
\begin{equation}
\begin{split}
 \mathrm{Re}(\varphi_k) &= \cosh(\gamma)\cos(2\pi\alpha k) - \cos(\pi\alpha),\\
\mathrm{Im}(\varphi_k) &= -\sinh(\gamma)\sin(2\pi\alpha k).
\end{split}
\label{eq:phi_decomp}
\end{equation}
Because $\beta\varphi_k$ for poles satisfies Eq.~\eqref{eq:poles} , the real part of $\varphi_k$ in Eq.~\eqref{eq:phi_decomp} must be zero, and the imaginary part of $\varphi_k$ satisfies $\beta\mathrm{Im}(\varphi_k) = (2n+1)\pi/2$, i.e.,
\begin{equation}
\left\{
\begin{array}{ll}
\cosh(\gamma)\cos(2\pi\alpha k) - \cos(\pi\alpha) = 0,\\ \\
-\beta\sinh(\gamma)\sin(2\pi\alpha k) = \frac{(2n+1)\pi}{2}.
\end{array}
\right.
\label{eq:betak}
\end{equation}
By solving Eq.~\eqref{eq:betak} and canceling $k$, we obtain the $n$-dependent $\beta$ for poles as
 \begin{equation}
     \beta=\frac{-1}{\sinh(\gamma) \sqrt{1 -[\cos(\pi\alpha)/\cosh(\gamma)]^2}}  \frac{(2n+1)\pi}{2}.
     \label{eq:beta}
 \end{equation}
Since the gain-loss-modulated localization enables the same cascade period of delocalization transitions as the period of poles in Eq.~\eqref{eq:beta}, we can thus express the cascade period $\Delta\beta$ associated with delocalization transitions as the difference between $\beta$ corresponding to indices $n$ and $n+1$:
\begin{equation}
  \Delta\beta = \frac{\pi}{\sinh(\gamma)}\frac{1}{\sqrt{1 - [\cos(\pi\alpha)/\cosh(\gamma)]^2}} .
  \label{eq:beta_period}
  \end{equation}

We now evaluate the profile of the complex IAAF potential $V_k(\beta)$ in the complex plane during one period of $\beta$. As an example, we choose a period range $[8.25 , 11]$ with the pole at $\beta \approx 9.63$ as the period center, $n = -4$, $\gamma=1$, and $\lambda = 1$, i.e., $\Delta\beta \approx 2.75$ from Eq.~\eqref{eq:beta_period}. We define $X=\mathrm{Re}(\phi_k)$, and $Y=\mathrm{Im}(\phi_k)$. According to Eq.~\eqref{eq:phi_decomp}, $X$ and $Y$ obey the following equation of an ellipse
\begin{equation}
\frac{\bigl(X + \cos(\pi\alpha)\bigr)^2}{\cosh^2(1)} + \frac{Y^2}{\sinh^2(1)} = 1.
\label{eq:ellipse}
\end{equation}
Using Eq.~\eqref{eq:potential}, we further express the real and imaginary parts of $V_k(\beta)$, namely $U(X,Y)$ and $W(X,Y)$, respectively:

\begin{equation}
\begin{split}
U(X,Y) = -\frac{1}{\tanh \beta}\frac{\sinh(2\beta X)}{\cosh(2\beta X) + \cos(2\beta Y)},\\
W(X,Y) = -\frac{1}{\tanh \beta}\frac{\sin(2\beta Y)}{\cosh(2\beta X) + \cos(2\beta Y)}.
\end{split}
\label{eq:transform}
\end{equation}

\begin{figure}
  \centering
  \includegraphics[width=1.0\linewidth]{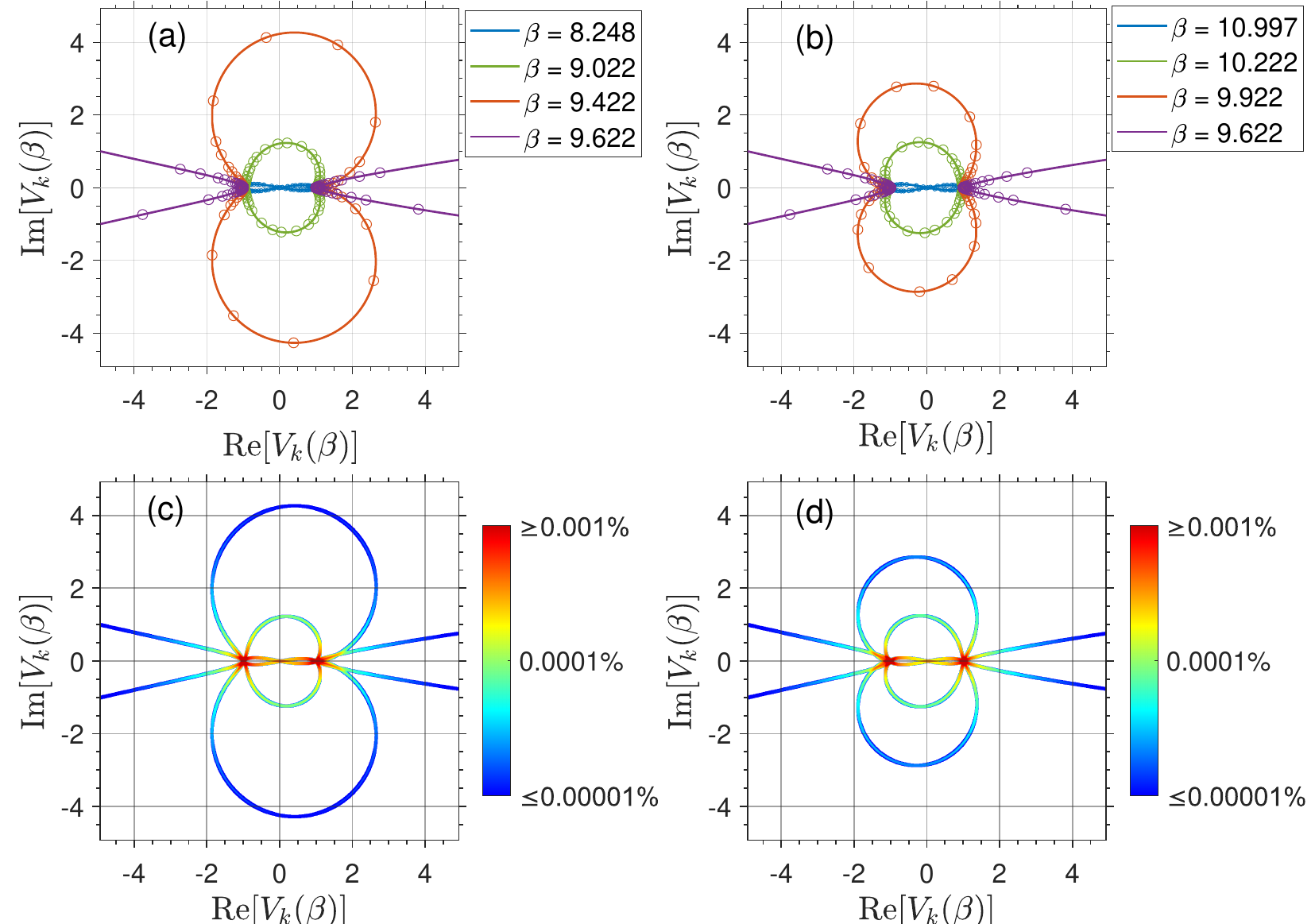}
   \caption{These curves show the profiles of $V_k(\beta)$ in the complex plane with $\gamma=1$, $\lambda=1$, and $k\in [1,  L]$ for (a) $\beta$ = 8.248, 9.022, 9.422, and 9.622, and (b) $\beta$ =9.622, 9.922, 10.222, and 10.997, where the circles mark the values of $V_k(\beta)$ for $L = 600$ as an example. (c) and (d) The corresponding probability plots of $V_k(\beta)$, where the colorbar represents the probability of lattice sites taking a given value of $V_k(\beta)$ for the total lattice size. Here $L = 85000$ is taken for curve profiles of $V_k(\beta)$.   }
  \label{fig:vk}
  \end{figure}

Figures ~\ref{fig:vk}(a) and~\ref{fig:vk}(b) show the profile of $V_k(\beta)$ in the complex plane $(U, W)$ with $\beta$ ranging from $8.25$ to $11$ during one period. Eqs.~\eqref{eq:ellipse} and~\eqref{eq:transform} indicate that the profile of $V_k(\beta)$ in the complex plane $(U, W)$ should be closed curves except the case of the poles, and thus the on-site potential of each site should lie on these curves. As seen in Figs.~\ref{fig:vk}(c) and~\ref{fig:vk}(d), the on‑site potentials of the majority of lattice sites concentrate near the real axis, whereas a subset of sites deviate significantly from the real axis and possess large imaginary or real parts. It is therefore inferred that these large imaginary potentials enable strong gain and loss, rendering localization even though the large real potential can slightly enhance localization.

\section{Periodic cascade of delocalization transitions}

To check the periodic cascade of delocalization transitions mainly produced by the competition between kinetic-energy-driven delocalization and gain-loss-modulated localization, we evaluate the inverse participation ratio (IPR) \cite{Evers2008} as a function of $\beta$. The IPR of the $n$-th eigenstate $\psi(E_n)$ with its eigenenergy $E_n$ is given by $\text{IPR}_n=\sum_{j=1}^{L}|\psi_j(E_n)|^4 \big/|\sum_{j=1}^{L}|\psi_j(E_n)|^2|^2$ where $L$ is the chain length. If the $n$-th eigenstate $\psi_n$ is extended, its IPR$_n$ is equal to the inverse of the system length, i.e., $\text{IPR}_n=1/L$, and drops to $0$ for an infinite system. Conversely, if $\psi_n$ is exponentially localized on $N$ sites, its IPR$_n$ is equal to $1/N$ and remains finite even for an infinite regime. However, to characterize the localization properties of a non-Hermitian system, the IPR$_n$ for $\psi_n$ within the biorthogonal form \cite{Brody2014} needs to be modified as
\begin{equation}
\mathrm{IPR}_n = 
\frac{
  \sum_{k=1}^{L} \left| \psi^L_{k,n} \, \psi^R_{k,n} \right|^2
}{
  \left| \sum_{k=1}^{L} \psi^L_{k,n} \, \psi^R_{k,n} \right|^2
}, 
\label{eq:ipr_npr}
\end{equation}
where the right and left eigenvectors satisfy
  \begin{equation}\label{eq:eigen}
  \begin{split}
     H |\psi_n^R\rangle = E_n |\psi_n^R\rangle,\\
  \langle\psi_n^L| H = E_n \langle\psi_n^L|,   
  \end{split}
  \end{equation}
with
\begin{equation}
\psi_m^{L\dagger} \psi_n^R = \delta_{mn}.
\label{eq:biorth}
\end{equation}
Unlike Hermitian system, we can't define a proper ground state \cite{Ashida2020}. We thus choose to evaluate the full-spectrum average IPR, i.e., $\langle\mathrm{IPR}\rangle=\frac{1}{L}\sum_{n=1}^{L}\mathrm{IPR}_n$ to characterize the global localization properties of the system. 

\begin{figure}
  \centering
  \includegraphics[width=0.95\linewidth]{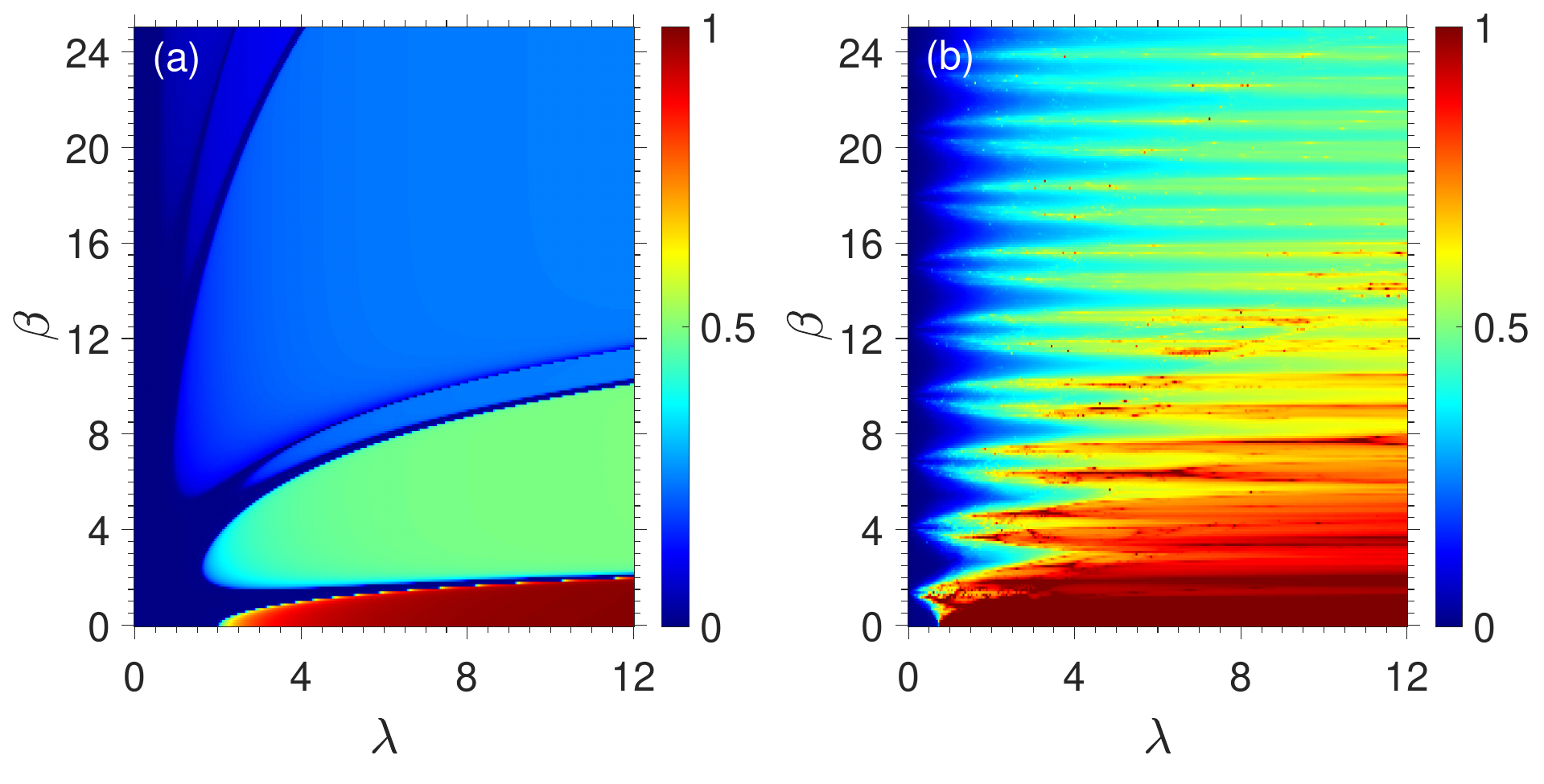}
   \caption{Phase diagrams in the ($\beta$, $\lambda$) plane for (a) the IPR of ground state of the Hermitian IAAF system with $\gamma=0$, and (b) the mean IPR of the non-Hermitian IAAF system with $\gamma=1$. Here the system size is $L=600$ for both cases. The emerged cascade phenomenon of delocalization transitions in (a) agrees with the earlier prediction \cite{Goblot2020}. A remarkable periodic cascade appears when $\lambda$ is less than about 4 in (b).  }
  \label{fig:ipr_comp}
  \end{figure}

Figure~\ref{fig:ipr_comp} shows the IPR of the ground state for Hermitian IAAF chain and the average IPR for non-Hermitian IAAF chain. The cascade phenomenon of delocalization transitions for the ground state of Hermitian IAAF chain emerges in Fig.~\ref{fig:ipr_comp}(a) and reproduces the earlier prediction \cite{Goblot2020}. When $\lambda$ is less than about 4, a remarkable periodic cascade of delocalization transitions emerges with respect to $\beta$ for the non-Hermitian system in Fig.~\ref{fig:ipr_comp}(b), compared with the irregular cascade of delocalization transitions in Hermitian system. If $\lambda$ is relatively strong, namely, $\lambda>4$, the potential energy dominates and thus the kinetic-energy-driven extended states vanish. Therefore, for a moderate potential magnitude the competition between kinetic-energy-driven delocalization and gain-loss-modulated localization renders a periodic cascade of delocalization transitions in the non-Hermitian IAAF model. 
 
\begin{figure}
  \centering
  \includegraphics[width=0.95\linewidth]{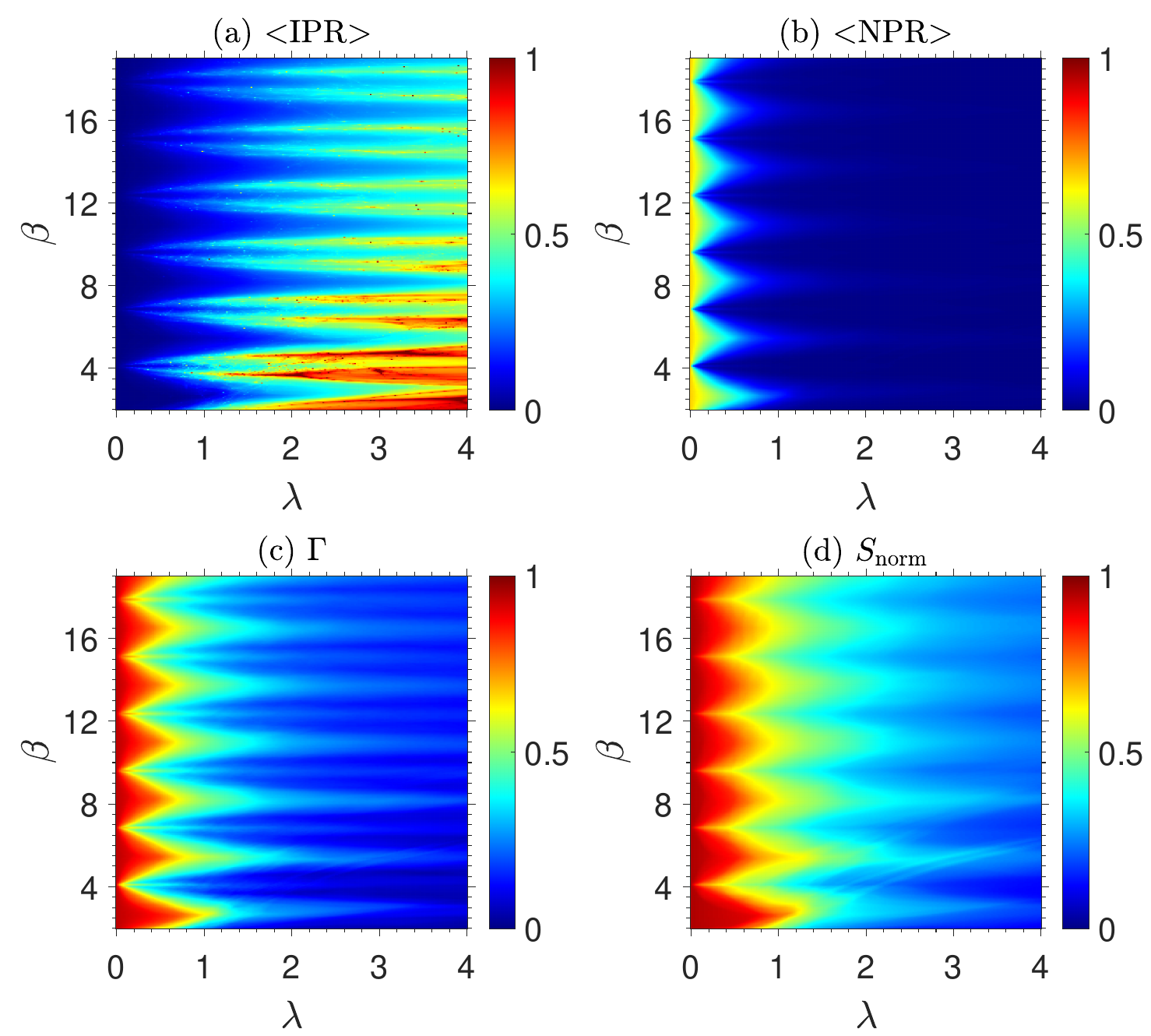}
   \caption{Phase diagrams of the non-Hermitian IAAF system in the ($\beta$, $\lambda$) plane with size $L=600$ and a fixed $\gamma=1$, for (a) $\langle\mathrm{IPR}\rangle$, (b) $\langle\mathrm{NPR}\rangle$, (c) $\Gamma$, and (d) $S_{\text{norm}}$. }
  \label{fig:four_color}
  \end{figure}

\begin{figure}
  \centering
  \includegraphics[width=0.95\linewidth]{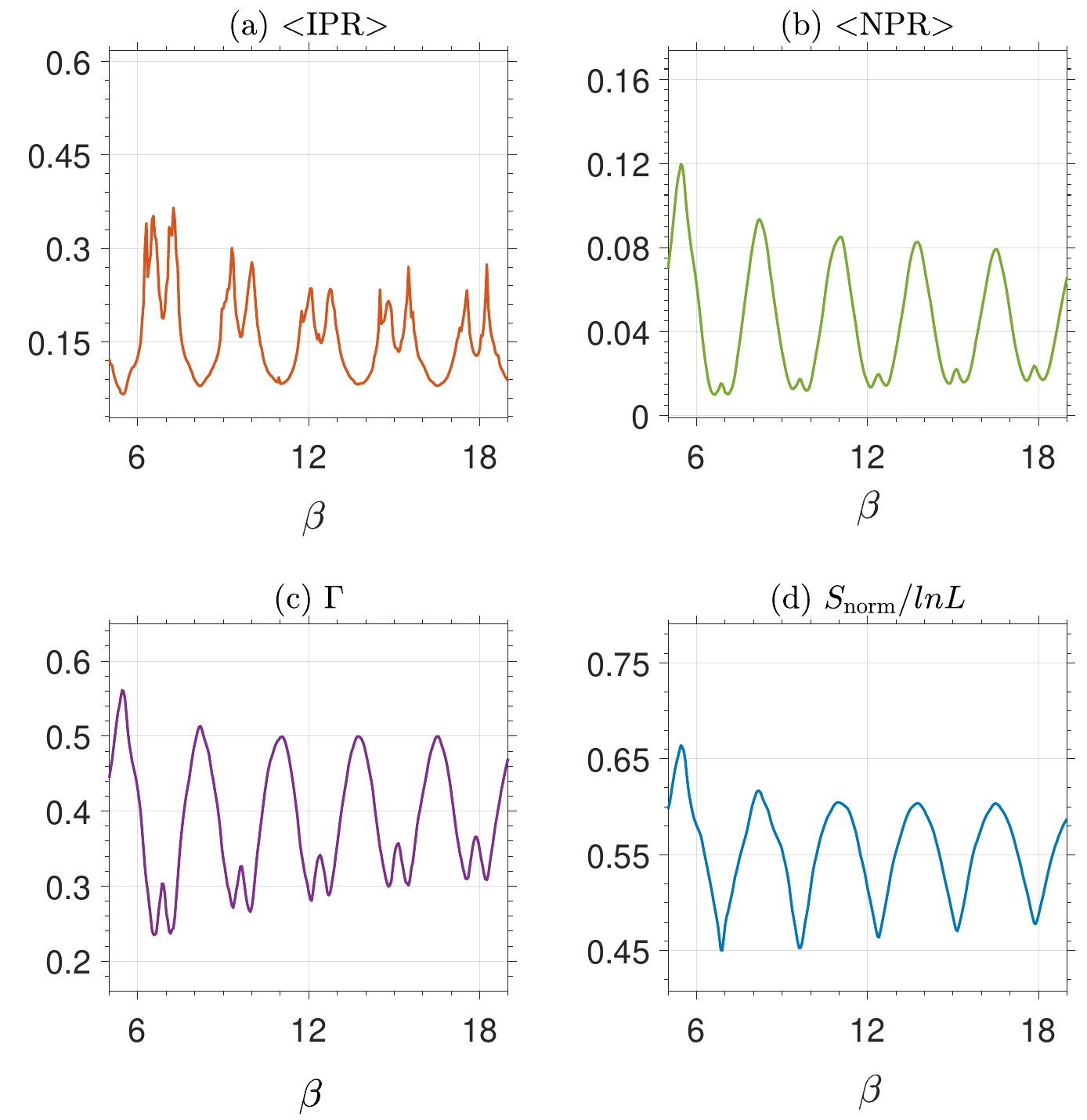}
   \caption{Corresponding line plots selected from Fig.~\ref{fig:four_color} with a fixed $\lambda = 1$, for (a) <IPR>, (b) <NPR>, (c) $\Gamma$, and  (d) $S_{\text{norm}}$. }
  \label{fig:four_line}
  \end{figure}

We also compute average normalized participation ratios $\langle\mathrm{NPR}\rangle$, fractal dimension $\Gamma$, and normalized Shannon entropy $S_{\text{norm}}$ \cite{Evers2008,Aditya2023_PRB107}, to further confirm the periodic cascade of delocalization transitions. These quantities have following corresponding definitions, 
\begin{equation}
\begin{gathered}
\langle\mathrm{NPR}\rangle = \frac{1}{L\,\langle\mathrm{IPR}\rangle},\\
\Gamma = -\frac{\ln\langle\mathrm{IPR}\rangle}{\ln L}, \\
S_{\text{norm}}=-\frac{1}{\ln L}\sum_{j=1}^{L}p_j\ln p_j,
\end{gathered}
\label{eq:Gamma}
\end{equation}
where $p_j=|\psi_j^L\psi_j^R|/\sum_{k=1}^{L}|\psi_k^L\psi_k^R|$. For extended and localized states with $L \to \infty$, $\langle\mathrm{NPR}\rangle$, $\Gamma$, and $S_{\text{norm}}$ take 1 and 0, respectively. Their numerical results are correspondingly plotted in Fig.~\ref{fig:four_color} with considered $\beta \in [2,19]$ and $\lambda \in [0,4]$. The remarkable periodic cascade of delocalization transitions occurs in the considered parameter regimes. For a fixed $\lambda$, for instance, $\lambda = 1$, $\langle\mathrm{IPR}\rangle$, $\langle\mathrm{NPR}\rangle$, $\Gamma$, and $S_{\text{norm}}$ do exhibit pronounced periodic variation with $\beta$, even though the oscillating magnitudes decrease slightly with the increased $\beta$ in Fig.~\ref{fig:four_line}. 

\begin{figure}
  \centering
  \includegraphics[width=1\linewidth]{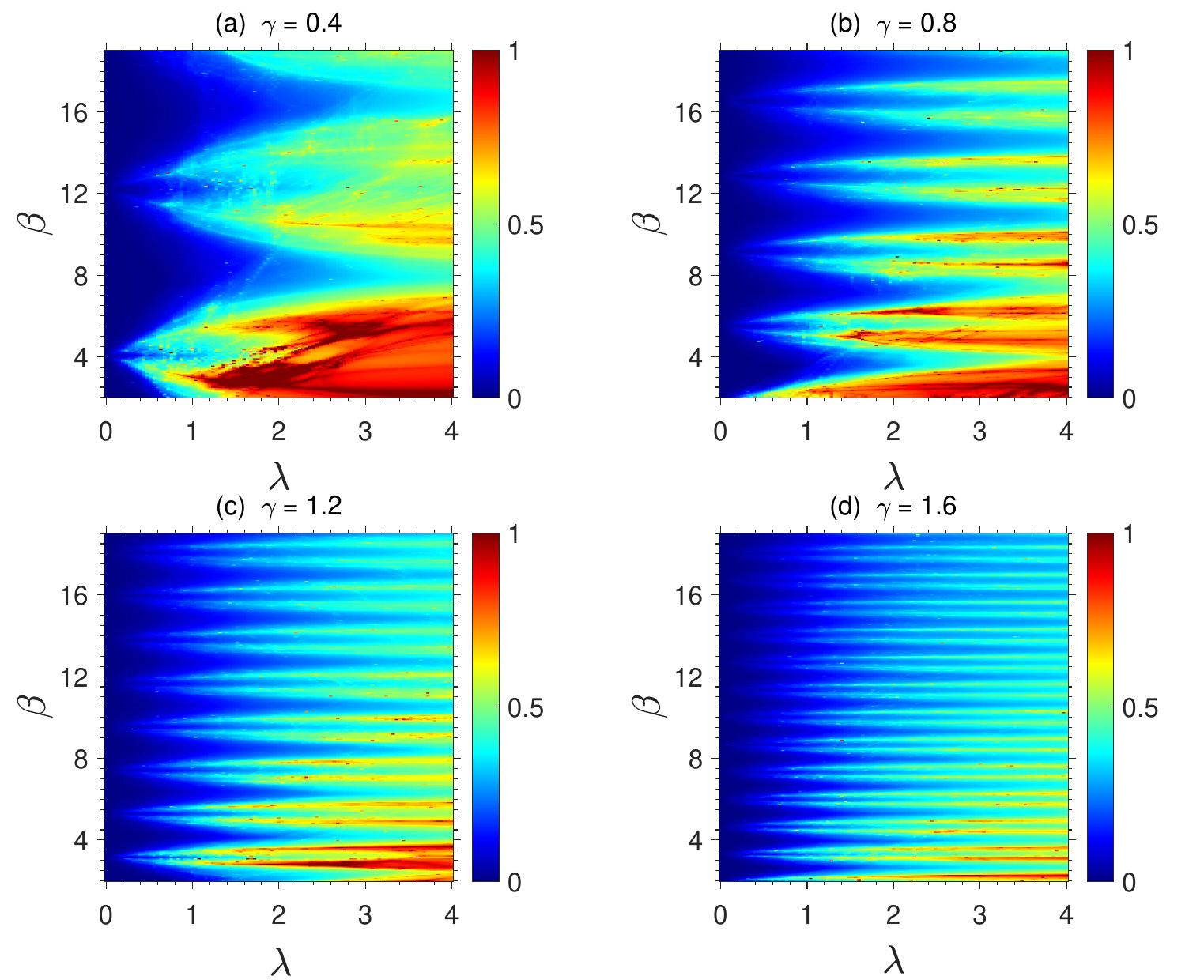}
   \caption{The $\langle\mathrm{IPR}\rangle$ of the non-Hermitian IAAF system with $\beta \in [2,19]$, and $\lambda \in [0,4]$, for $\gamma=$ 0.4, 0.8, 1.2, and 1.6, in (a)-(d), respectively. }
  \label{fig:four_gamma}
  \end{figure}
From Eq.~\eqref{eq:beta_period}, through changing $\gamma$ we can tune the cascade period $\Delta\beta$ of delocalization transitions. Fig.~\ref{fig:four_gamma} shows the $\langle\mathrm{IPR}\rangle$ of four different $\gamma$ including 0.4, 0.8, 1.2, and 1.6, respectively. As expected, different $\gamma$ renders different $\Delta\beta$, and the numerical results of $\Delta\beta$ agree with the analytical results of Eq.~\eqref{eq:beta_period}, i.e., $\Delta\beta=$ 8.12, 3.67, 2.12, and 1.34, respectively.

\section{Expansion dynamics}
In this section, we explore the tunable periodic cascade of delocalization transitions from the perspective of expansion dynamics. We set an initial state with its initial position at the chain center, i.e., $\psi_\mathrm{0}(k) = \delta_{k, L/2}$. By using biorthogonal eigenvectors \cite{Brody2014}, we can write the time-evolved wave function as $\psi(t)=\sum_n(\psi_n^{L\dagger}\psi_0)\, e^{-iE_n t}\, \psi_n^R$, where $n$ is the energy index. The root-mean-square
displacement $\sigma(t)$ describes the weight-averaged departure of a particle from its starting position as time evolves, and has been experimentally observed \cite{Billy2008,Roati2008,Lahini2009,Schreiber2011,Weidemann2021}. Therefore, $\sigma(t)$ is widely applied to characterize the dynamical behavior of the system. The $\sigma(t)$ has following definition
\begin{equation}
\sigma(t) = \sqrt{\sum_{i=1}^{L} (i - i_0)^2\, |\tilde{\psi}_i(t)|^2},
\label{eq:msd}
\end{equation}
where $|\tilde{\psi}(t)\rangle = |\psi(t)\rangle / \sqrt{|\langle\psi(t)|\psi(t)\rangle|}$, and $i_0 = L/2$ is the initial position. Note that in numerical calculations we have shifted the energy eigenvalues $E_n$ as $E_n-i\max[\mathrm{Im}(E_n)]$ to avoid divergence. In addition, the energy shift does not change $\sigma(t)$.

For extended states, $\sigma(t)$ grows swiftly over time before reaching saturation. By contrast, $\sigma(t)$ hardly evolves for localized states. In the critical-regime systems, $\sigma(t)$ still rises as time elapses yet exhibits a slower growth rate relative to that of extended states. For non-Hermitian system, it can be inferred that the system will exhibit different behaviors in short and long time cases since the system will eventually evolve to eigenstates with the largest imaginary part \cite{Ashida2020,Longhi2022,Spring2024}. Therefore, we analyze delocalization transitions by considering both short-time and long-time $\sigma(t)$ cases.

\begin{figure}
  \centering
  \includegraphics[width=0.95\linewidth]{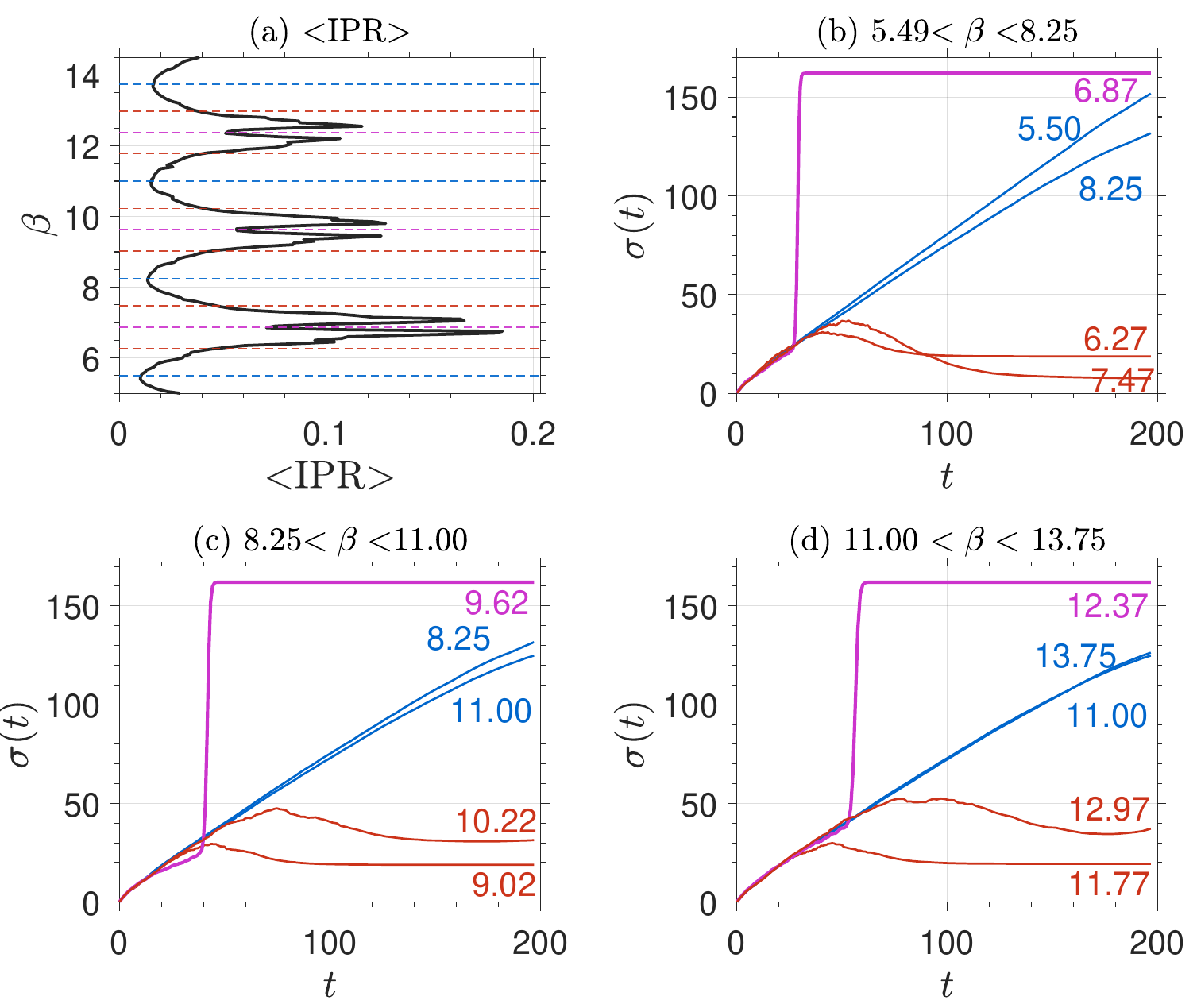}
   \caption{(a) The same <IPR> as a function of $\beta$ as that in Fig.~\ref{fig:four_line}(a), where some chosen values of $\beta$ are picked up by dashed lines. (b)-(d) Short-time $\sigma(t)$ for different $\beta$ chosen in (a). }
  \label{fig:sigma_low}
  \end{figure}

Figure.~\ref{fig:sigma_low}(a) presents the $\langle\mathrm{IPR}\rangle$ as a function of $\beta$ with $\lambda=1$, exhibiting the periodic cascade of delocalization transitions. Figs.~\ref{fig:sigma_low}(b)-\ref{fig:sigma_low}(d) show short-time $\sigma(t)$ of system with corresponding different $\beta$ denoted by the dashed lines in Fig.~\ref{fig:sigma_low}(a), in three continuous cascade periods of delocalization transitions, i.e., for $\beta=$5.50, 6.27, 6.87, 7.47, and 8.25 in period 1, $\beta=$8.25, 9.02, 9.62, 10.22, and 11.00 in period 2, and $\beta=$11.00, 11.77, 12.37, 12.97, and 13.75 in period 3. For $\beta=$5.50, 8.25, 11.00, and 13.75, $\sigma(t)$ expands over time, reflecting the extended state behavior. For $\beta=$6.27, 7.47, 9.02, 10.22, 11.77, and 12.97, $\sigma(t)$ initially expands from the delta-function initial state and then almost remains unchanged over time, indicating the feature of localized states. These short-time behaviors for the three continuous $\beta$ periods verify the periodic cascade of delocalization transitions in the non-Hermitian IAAF system. We also see that for $\beta=$6.87, 9.62 or 12.37, $\sigma(t)$ jumps quickly over a short time and immediately stays at the sites with the divergent imaginary potential at the pole condition. Note that this localization resulting from the divergent imaginary potential is corresponding to pseudo-localization\cite{Longhi2022}, because the divergent imaginary potential enables the wave function to be immediately pinned at these sites.

\begin{figure}
  \centering
  \includegraphics[width=1\linewidth]{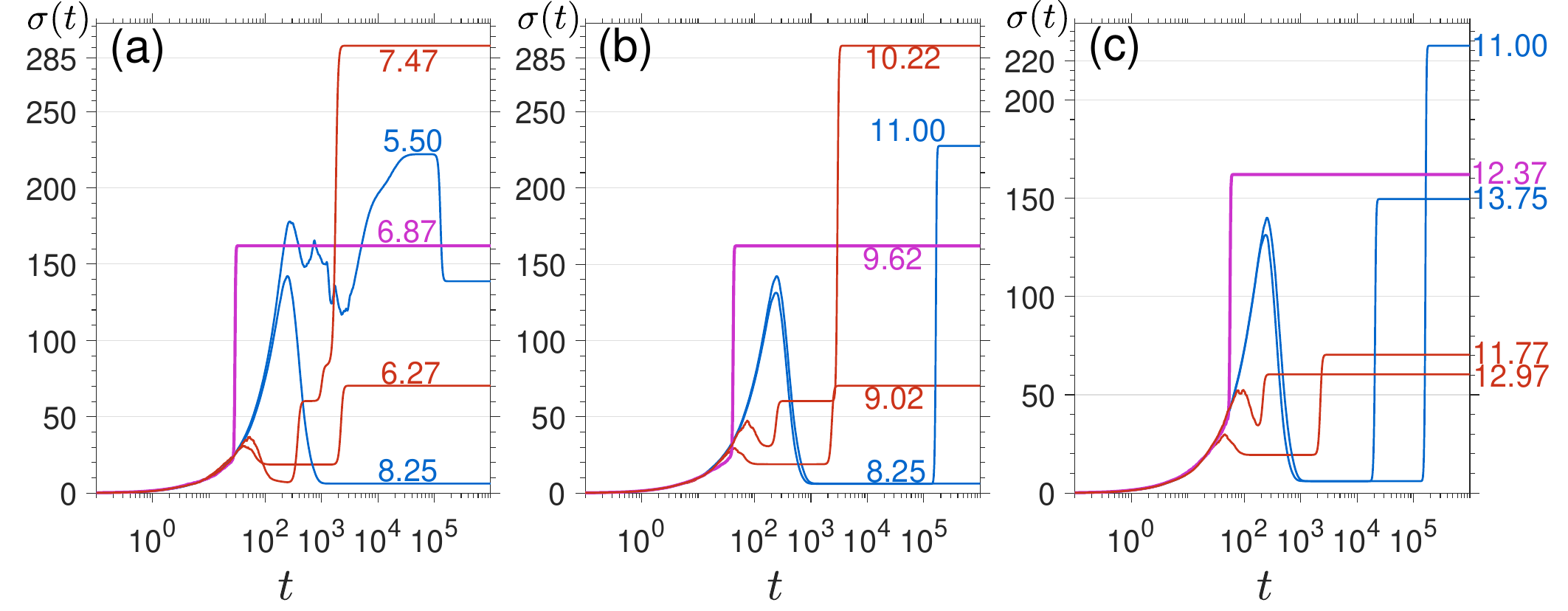}
   \caption{(a)-(c) Long-time $\sigma(t)$ for corresponding $\beta$ given in Fig.~\ref{fig:sigma_low}(a). }
  \label{fig:sigma_long}
  \end{figure}

Figure.~\ref{fig:sigma_long} shows long-time $\sigma(t)$ of system with the same $\lambda$ and $\beta$ in Fig.~\ref{fig:sigma_low}. With increased time, for extended states, instead of continuous rising, their $\sigma(t)$ quickly drop or undergo fluctuations, and finally jump to some fixed sites and keep unchanged. For localized states, their $\sigma(t)$ also finally jump to some fixed sites and stay. On sufficiently long time scales, all complex-energy states, whether extended or localized, jump to some fixed sites. Such behavior is called  pseudo-localization \cite{Longhi2022}, which arises from a purely dynamical gain‑selection effect, i.e., complex-energy states $\ket{\psi_n}$ with complex energy $E_n$ eventually grow exponentially according to $e^{-iHt}\ket{\psi_n} = e^{-i\mathrm{Re}(E_n)t}\,e^{\mathrm{Im}(E_n)\,t}\,\ket{\psi_n}$.

\begin{figure}
  \centering
  \includegraphics[width=0.95\linewidth]{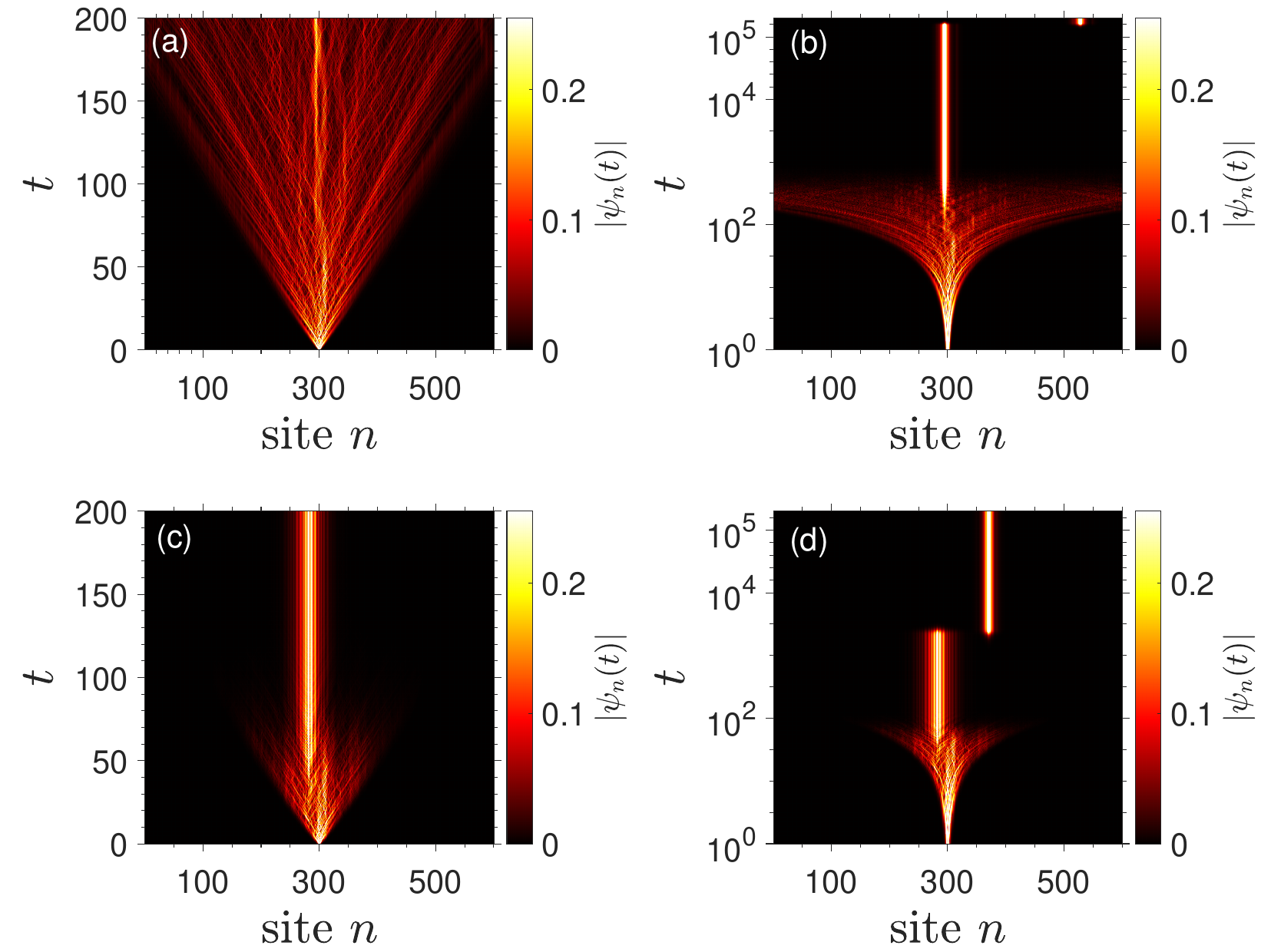}
   \caption{Wave spreading profiles in the one-dimensional non-Hermitian IAAF chain, for short-time and long-time cases in (a) and (b) with $\beta=8.25$, respectively, and for short-time and long-time cases in (c) and (d) with $\beta=10.23$, respectively. Here, $\lambda=1$, $\gamma=1$, and $L=600$.  }
  \label{fig:spread}
  \end{figure}
 
To better visualizing the localized and extended states within the periodic cascade of delocalization transitions at the short-time dynamics as well as the pseudo-localization states at the long-time dynamics, we plot the short-time and long-time wave spreading profiles for the extended state with $\beta = 8.25$ in Figs.~\ref{fig:spread}(a) and ~\ref{fig:spread}(b), and the localized state with $\beta = 10.23$ in Figs.~\ref{fig:spread}(c) and ~\ref{fig:spread}(d), respectively. As shown in Fig.~\ref{fig:spread}(a), at short times, the wave function quickly spreads over the whole chain for the extended state; for the localized state in Fig.~\ref{fig:spread}(c), after an initial spreading, the wave function contracts and settles into a localized profile confined to only a few lattice sites. After sufficiently long time scales, both the extended and localized states hop to a few fixed sites, exhibiting pseudo-localization characteristics, as shown in Figs.~\ref{fig:spread}(b) and ~\ref{fig:spread}(d). Even though the pseudo-localization has the same period, it is not our central result. The periodic cascade of delocalization transitions resulting from the competition between kinetic-energy and finite imaginary potential is our central finding.

\section{Conclusions}
In summary, we revealed that, in the non-Hermitian IAAF model with an imaginary phase shift under a moderate potential magnitude, the competition between kinetic-energy-driven delocalization and gain-loss-modulated localization gives rise to a periodic cascade of delocalization transitions. We derived the analytical expression of the cascade period of delocalization transitions, which is only determined by the imaginary phase shift characterizing the non-Hermiticity. The analytical expression indicates that the larger the imaginary phase shift is, the smaller the cascade period is. We performed numerical calculations on mean IPR, mean NPR, fractal dimension, and normalized Shannon entropy of the non-Hermitian IAAF model, and our numerical results confirm the periodic cascade phenomenon of delocalization transitions as well as theoretical predictions. 

Besides the static localization analysis, we also explored short-time and long-time dynamical behaviors of wave-packet expansion during delocalization transitions via the time-dependent root-mean-square displacement and wave spreading profiles. The calculated short-time dynamical results verify the periodic cascade of delocalization transitions in the non-Hermitian IAAF system. In addition, pseudo-localizations of wave-packet expansions appear at the pole condition. On sufficiently long time scales, it is found that all complex-energy states eventually evolve into  pseudo-localization states owing to the exponential growth of wave functions for complex energy states. For larger potential magnitude, the kinetic-energy-driven extended states vanish and hence localization dominates. Our finding gives a new type of  cascaded delocalization transitions with a tunable cascade period and opens a route toward engineering periodic cascades of delocalization transitions in experiments.

\section{Acknowledgments}
This work is supported by the National Natural Science Foundation of China (Grant No. 12247101), the National Key Research and Development Program of China (Grant No. 2022YFA1402704), the Fundamental Research Funds for the Central Universities (Grant No. lzujbky-2025-jdzx07), the Natural Science Foundation of Gansu Province (No. 22JR5RA389, No.25JRRA799), and the 111 Center under Grant No. B20063.
\bibliography{bibfile}
\end{document}